# Mental states follow quantum mechanics during perception and cognition of ambiguous figures.

Elio Conte[1,2], Andrei Yuri Khrennikov[3], Orlando Todarello[4], Antonio Federici[1], Leonardo Mendolicchio[5], Joseph P. Zbilut [6]

1. Department of Pharmacology and Human Physiology – TIRES – Center for Innovative Technologies for Signal Detection and Processing, University of Bari- Italy;
2. School of Advanced International Studies for Applied Theoretical and Non Linear Methodologies of Physics, Bari, Italy;
3. International Center for Mathematical Modeling in Physics and Cognitive Sciences, MSI, University of Växjö, S-35195, Sweden;
4. Department of Neurological and Psychiatric Sciences, University of Bari, Italy;
5. Department of Mental Health, University of Foggia, Italy;
6. Department of Molecular Biophysics and Physiology, Rush University Medical Center, 1653 W, Congress, Chicago, IL 60612, USA.

Correspondence to: Elio Conte   (email: elio.conte@fastwebnet.it)
tel. and fax (39) 080 5478414

Abstract:
Processes undergoing quantum mechanics, exhibit quantum interference effects. In this case quantum probabilities result to be different from classical probabilities because they contain an additional main point that in fact is called the quantum interference term. We use ambiguous figures to analyse if during perception cognition of human subjects we have violation of the classical probability field and quantum interference. The experiments, conducted on a group of 256 subjects, evidence that we have such quantum effect. Therefore, mental states, during perception cognition of ambiguous figures, follow quantum mechanics.

43
Introduction

Mental operations consists of a content plus the awareness of such content. Consciousness is a system which observes itself. It evaluates itself being aware at the same time of doing so. We may indicate awareness statements by a, b, c, … that are self-referential or auto referential and content statements of our experience by x , y, z, … . a=F(a, x) is the most simple definition of a single autoreferential statement. For example, x= the snow is white; a=I am aware of this. Human experience unceasingly involves intrinsically mental and experiential functions such as "knowing" and "feeling", involving images, intentions, thoughts and beliefs. A continuous interface holds between mind/consciousness and brain. Neuroscience and neuro-psychology have reached high levels of understanding and knowledge in this field by the extended utilization of electrophysiological and of functional brain imaging technology. First of all this last technique has identified brain areas that are involved in a wide variety of brain functions including learning and memory. These are valuable studies that provide knowledge of the functional role of different brain areas. However neuroscience finds it hard to identify the crucial link existing between empirical studies that are currently described in psychological terms and the data that arise instead described in neurophysiological terms. It is assumed that the measurable properties of the brain through functional imaging technology should be in itself sufficient to achieve an adequate explanation of the psychologically described phenomenology that occurs during neuropsychological experiments. Of course, this manner of investigation encounters the reservation of some investigators who suggest that intrinsically mental and experiential functions such as "feeling" and " knowing" cannot be described exclusively in terms of material structure, and they require an adequate physics in order to be actually explained. To this purpose they outline the important role that quantum mechanics could carry out. In particular, we outline here the effort of Stapp in several years and more recently (Schwartz, Stapp, Beauregard, 2005), who repeatedly outlined the problem to consider quantum mechanics in Neuroscience and Neuropsychology. The prospects for a quantum neurobiology were also outlined already more than a past decade ago (Miller 1996). Therefore, it becomes of fundamental and general interest for neuroscience and neuro-psychology to ascertain by experiments if quantum mechanics has a role in brain dynamics. In the present paper we present an important contribution concerning this basic problem. We demonstrate that mental states follow quantum mechanics during perception and cognition of ambiguous figures.

Quantum theoretical approach

Previously we have given a logical self-reference mathematical model of conscious experience that is due to A.G. Kromov (Khromov 2001). Consciousness represents the hard problem for scientific, epistemological and philosophical knowledge (Whitehead 1929; Whitehead 1933; Shimony 1997; Stapp 1933). Present physical theory does not define an apparatus to describe conscious systems. However, we cannot exclude that future generalizations of the present physical knowledge will be able to approach such basic problem. An indication arises from quantum mechanics. Quantum theory represents the most confirmed and celebrated theory of science. Started in 1927 by founder fathers as Bohr, Heisenberg, Schrödinger, and Pauli, it has revolutionized our understanding of the physical reality in both scientific and epistemological fields. It was introduced to describe the behaviour of atomic systems but subsequently its range of validity has turned out to be much wider including in particular some macroscopic phenomena like superconductivity or superfluidity. There is a salient and crucial feature for this theory. The conceptual structure and the axiomatic foundations of quantum theory repeatedly suggested from its advent and in the further eighty years of its elaboration that it has a profound link with mental entities and their dynamics. From its advent such theory was strongly debated but often also criticized just for its attitude to prospect a model of reality that results strongly linked to mental entities and their dynamics. The standard formulation of quantum mechanics seems to fix the necessity for to admitting the unequivocal role of mental properties to represent properties of the physical objects. We retain that it represents an important feature of the theory instead of its limit. However, there is the problem to correctly interpreting the connection between quantum mechanics and mental properties in the sphere of reality. It must be clear that one cannot have in mind a quantum physical reduction of mental processes. N. Bohr (Bohr 1987) borrowed the principle of complementarity, which is at the basis of quantum mechanics, from psychology. He was profoundly influenced from reading the "Principles of Psychology " by W. James (James 1890).  However, N. Bohr never had in mind quantum-reductionism of mental entities. Starting with 1930, there was also an important correspondence between W. Pauli and C.G. Jung that culminated in the formulation of a theory of mind-matter synchronization (Meier 2001). Also  in this case these founding fathers as Pauli and Jung were very distant to consider a quantum-reductionism perspective. V. Orlov (Orlov 1982) proposed a quantum logic to describe brain function but also he did not look for reduction of mental processes to quantum physics. The correct way to frame the problem is not to attempt a quantum reduction of mental processes. The most profitable applications of quantum mechanics in cognitive sciences and psychology can be obtained not by any attempt of quantum physical reduction but  giving experimental evidence that cognitive systems are  very complex information systems, to which also some laws of quantum systems can be applied. Just the reaching of such objective would represent a very great advance in the domain of knowledge. In fact, starting with such experimental evidence, we could elaborate some future developments knowing this time the principles to use, the  formal criteria to follow in order to approach with higher  rigour the framing of the nature of mental entities and of their dynamics.

We retain that in this perspective we obtain here a first contribution since we give  for the first time experimental confirmation that mental states, at some stages of human perception and cognition, can be described by the formalism of quantum mechanics. Thus for the first time, also if not under a reductionism perspective, we have the chance to understand what are the principles and rules acting as counter part of human mind. To fully agree with the present paper, the reader must take care the following crucial point: quantum mechanics has its unique law of transformation of probability distribution. It is well known that the main feature of quantum probabilistic behaviour is the well known phenomenon of interference of probabilities. Such interference regime may be obtained only in quantum systems, e. g., in the celebrated two slit experiment that has been confirmed at any level of experimental investigation (Feynman and Hibbs 1965; Zelinger 1966; Ballentine 1970). Quantum dynamics of human decision making was also studied rather recently by Busemeyer et al. (Busemeyer et al 2006). The interference gives the experimental basis of the superposition principle and this latter is the basis foundation of the physical and philosophical system of view that we call quantum mechanics. This is the essential peculiarity that we aim to investigate in the present paper. Recently, the problem of quantum probabilities was extended to the so called calculus of contextual probabilities (Khrennikov 2001-2006). The basic notion of this approach is the *context*. In quantum mechanics it is a complex of experimental physical conditions. In the present paper it will be a complex of mental conditions. The essential feature of this elaboration is that by it we may be able to ascertain the presence of quantum like behaviour also in systems that exhibit context quantum like behaviour as physical, cognitive, social systems. We will not enter in the detail of the method here for brevity but all the features are given in the quoted literature (Khrennikov 2001-2006). The essence of the method is based on the following step. Let A and B be two dichotomous questions which can be asked of people, S, with possible answers "yes (+) or not (-)". In our case we consider A and B two mental quantum like observables of people S under investigation. We split the given ensemble S of humans into two sub ensembles U and V of equal numbers. To ensemble U we pose the question A with probability in answering, given respectively by $p(A=+)$ and $p(A=-)$, and $p(A=+)+p(A=-)=1$. We pose the question B immediately followed by the question A to the ensemble V. We calculate conditional probabilities $p(A=+/B=+)$ and $p(A=+/B=-)$ and equivalent probabilities for the case ($A=-$). We reach in this manner a no evadable feature of such experiment.

Let us recall the fundamental law of classical probability theory, the Bayes formula of total probability (FTP):

$$p(A=+) = p(B=+)p(A=+/B=+) + (B=-)p(A=+/B=-)$$

It plays a fundamental role in classical statistics and decision making. However, it is violated for statistical experiments with quantum systems. The two slit (interference) experiment is the basic experiment violating FTP. In physical literature such a viewpoint on this experiment was presented in detail by Feynman (Feynman, Hibbs 1965). By itself the appearance of interference fringes was not surprising for him: in principle, interaction with a screen with slits may produce any kind of distribution of points on the registration screen. Quantum probabilistic features appear if one consider three different experiments: a) only the first slit open (in an equivalent manner we consider here the case B=+1), b) only the second slit is open (we consider in this case B= -1), c) both slits are open. Here the random variable B determines the slit. We now choose any point at the registration screen. The random variable A is A= +1 if a particle hits the screen at this point and A=-1 in the opposite case. For classical particles, FTP should predict the probability $p(A=+)$ for the c-experiment (both slits are open) on the basis of probabilities $p(A=+/B=+)$, $p(A=+/B=-)$ which are provided by the a and b- experiments. But, as was already mentioned, FTP is violated for quantum particles: an additional cosine-type term appears in the right-hand side of FTP. This is nothing else than the quantum interference of probabilities. Feynman characterized this feature of the two slit experiment as the most profound violation of laws of classical probability theory. Our aim is to show that this fundamental law of classical probability can be violated even by cognitive systems. Opposite to quantum mechanics, we could not start directly with the Hilbert space formalism. In quantum mechanics this formalism was justified by experiments. It was not yet done in cognitive science and psychology. Here we have to start directly with experimental data, and we calculate the quantity representing deviation from the classical probabilistic law (Khrennikov 2001-2006). It represents quantum interference and it is given in the following terms:

$$\lambda = \frac{p(A=+) - p(B=+)p(A=+/B=+) - p(B=-)p(A=+/B=-)}{2\sqrt{p(B=+)p(A=+/B=+)p(B=-)p(A=+/B=-)}} = $$
$$= \frac{\Delta p}{2\sqrt{p(B=+)p(A=+/B=+)p(B=-)p(A=+/B=-)}}$$
(2.1)

We recall that the conventional quantum formalism implies that in general $\lambda$ is not equal to zero (opposite to classical statistics). If in the experiment for cognitive systems with questions A and B results $\lambda \neq 0$ it will be certain that we are in presence of quantum like behaviour for mental states owing to the presence of interference terms for the calculated probabilities. In the case $\lambda = 0$ we will conclude that quantum like behaviour is absent in the dynamic regime of our mental states. We also recall that in quantum mechanics the coefficient $\lambda$ can be represented as $\lambda = \cos \vartheta$, where $\vartheta$ is obviously a well known angle of phase. We may expect a similar result for cognitive systems. In conclusion, the interference coefficient, introduced in the (2.1), gives a measure of incompatibility of different contexts.

We may also proceed giving a quantum like framework of mental states. Let us remember that, according to Born's probability rule (Bohr 1987; Khrennikov 2001-2006), we have

$$P(A = \pm) = |\varphi(\pm)|^2 \qquad (2.2)$$

In the case in which the experiment confirms quantum mechanics in dynamics of mental states, as in standard theory, we can write a quantum-like wave function $\varphi_S(\pm)$ relative to the mental state $S$ of the population investigated, and it will be represented by the complex amplitude as for the first time elaborated in and applied in our previous papers (Khrennikov 2001-2006):

$$\varphi_S(x) = [P(B = +)P(A = x/B = +)]^{1/2} + e^{i\vartheta(x)} [P(B = -)P(A = x/B = -)]^{1/2} \text{ with } x = \pm \qquad (2.3)$$

It is necessary to outline here the importance of future studies on cognition based possibly on the (2.3). To this purpose, we would add something of more specific in relation to the meaning of the wave function given in the (2.3). The image recognition could be characterized by synchronization of firings in a neural network responsible for image recognition. Such a synchronization may be conceived as a stabilization to a fixed frequency of firings, and thus can be considered as a version of the collapse of the wave function. In substance, before synchronization-collapse, the quantum like state of the particular neural network that we are considering, is still characterized by superposition of frequencies of neural firing. Actually, it may be interpreted as superposition of alternatives.

In our case these alternatives are two ambiguous sub-pictures, say A(+) and A(-), in a given A-figure. The quantum like state of the neural network working with image recognition is in the superposition of two states, say $\varphi(A+)+\varphi(A-)$. In the ideal case this superposition is induced by superposition of neural oscillations on two definite frequencies. But in reality, of course, each state $\varphi(A+)$ and $\varphi(A-)$ is realized at neural level by its own range of frequencies.

Arrangement of the Experiment

The experiment was based on the search of quantum behaviour in mental states during human perception and cognition of ambiguous figures. We used ambiguous figures in this paper not to analyse the field of the optical illusions but to consider the perceptual-cognitive system in a simple model but in the perspective of possible analysis in future of more complex conditions of perception and cognition.

In general, it is known that the brain organizes sensory input into some representation of a given environment. Studies of perception indicate that the mental representation of a visually perceived object at any instant is unique even if we may be aware of the possible ambiguity of any given representation. The well known example is the Necker cube (Necker 1832) where we see the cube in one of two ways and only one of such representations is apparent at any time. Therefore the basic key of the experiment that recalls a quantum possible behaviour, is in the following statements. We may be able to see the ambiguity of the design and even we may be able to switch wilfully between representations: we can be aware that multiple representations are possible but we can perceive them only one at time, that is serially. Let us see the quantum like model that arises from this statement.

Bistable perception is induced whenever a stimulus can be thought in two different alternatives ways. Previously (Conte et al. 2003; Conte et al. 2007), we proposed to describe bistable perception with the formalism of a two quantum system. In our quantum like model of mental states we consider that an individual can potentially have multiple representations of a given choice situation, but can attend to only one representation at any given time. In this quantum mechanical framework we distinguish a potential and an actual or manifest state of consciousness. The state of the potential consciousness will be represented by a vector in Hilbert space. If we indicate for example a bi dimensional case with potential states $|1>$ and $|2>$, the potential state of consciousness will be given by their superposition

$$\psi = a|1> + b|2>.  \qquad (3.1)$$

Here, $a$ and $b$ represent probability amplitudes so that $|a|^2$ will give the probability that the state of consciousness, represented by percept $|1>$, will be finally actualised or manifested during perception. Conversely $|b|^2$ will represent the probability that state(percept) $|2>$ of consciousness will be actualised or manifested during perception. It will be $|a|^2 + |b|^2 = 1$.

In a quantum mechanical model of consciousness, as outlined from various authors and in particular by Manousakis (Manoussakis 2006; Atmanspacker et al. 2004), we admit that, when a conscious observation happens, the actual perceptual event that in correspondence is realized in consciousness, is linked to a particular neural correlate brain state. In this manner, in the (3.1), $|1>$ and $|2>$ represent two possible states having two distinct neural correlate of consciousness in brain states.

For brevity we will not consider here the case of the evolution in time of the state of potential consciousness.

In conclusion, according also to Eccles and Beck (Beck, Eccles 1992), the mind is a field of probability. Quantum mechanics seems to relate tightly our mental entities. The (3.1) ,given by the superposition principle, links the condition of doubt , or of inner conflict , or still of inner indetermination and uncertainty that we may have in front of a novel question posed to our perception and cognition.

Experiment Set Up.

Images that can be perceived in at least two mutually exclusive manners, define what we call a multistable perception. In it the physical input to the retina remains constant but perceptual interpretations of the ambiguous image alternate between the percept possibilities. Generally speaking, the problem is to explain how, given multiple possibilities of representation, a particular representation can take place over our attention. In the case of Necker cube transitions between percepts, it may be possibly stochastic, but in more complex mental and psychological situations some underlying factors may give the edge to one representation over another. Recent or repeated prior use of a representation may play role in advantaging one representation over the other. This is the reason to project the experiment carefully. Otherwise, the study of ambiguous figures has intrigued and still is of valuable interest for psychologists and neuroscientists. A variety of theories has been published (Kohler 1940; Gibson, 1950; Atteneve 1971; Turvey, 1977; Toppino, Long 1987; Horlitz, O'Leary 1993; Mitroff et al. 2005). For brevity, we will mention here the so called "low level" and "high level" approaches. The first evidences that reversals are due to adaptation to feedforward mechanisms, the second instead supports that reversals arise in a feedback way on lower level sensory mechanisms. For the purposes of our experimentation we evaluated that two types of observers exist : 1) fast observers, having larger frequency of perspective reversals and 2) slow observers whose frequency is lower. The time of staying one of the two percepts are on average on the order of two seconds, but may also approach about five seconds. To further confirm our quantum model with potential and actual states of consciousness, we have a further phenomenological datum. Subjects demonstrate uncertain time in percept states in addition to perspective reversal. Uncertain times about 1 sec were experimentally ascertained on average for fast subjects (Sakai et al. 1993; Sakai et al. 1994). In conclusion, two kinds of times are identified during experiments with ambiguous figures: a time persisting one of the two possible percepts that may be called the time persisting for percept, and a time of uncertainty for which neither of the percepts is certain for the subject, and it mirrors the previous quantum model on potential state of consciousness. There are still two basic different approaches in studies of perception of ambiguous stimuli. One is the behavioural response to a stimulus based on psychological or mental processes. This is usually investigated using the frequency of reversals. The second approach looks instead to neural correlates of psychological processes triggered by stimuli. Neurophysiological and neuropsychological investigations have started to identify the physiology of such percept reversals. Recent fMRI studies have suggested that conscious detection of visual changes relies on both parietal and frontal areas (Kleinschmidt et al. 1998; Inui et al. 2000). These areas, therefore, seem to play an important role in detecting changes in our perception, whether they are caused externally or internally. Electrophysiological studies have been developed evaluating changes in neuronal activity related to perceptual reversals. In particular, ERP studies identified a P300-like component related to perceptual reversals (Struber et al. 2000; Struber, Herrmann. 2002; Pitts et al. 2007).

Methods

Our experiment was based on the analysis of (2.1) with ambiguous figures A and B. Previously, we performed four experiments of this kind based on ninety eights subjects (Conte et al. 2003; Conte et al. 2007). We outline here that also maintaining the same methodology of the previous paper, we performed a completely new experimentation. We analysed a group of 72 subjects giving geometrical figures as Test A and Test B, respectively. See Figures 1 here attached. Still, we analysed a group of 52 subjects giving this time ambiguous figures of animals as Test A and Test B respectively, (see still figures 2 for the test) We performed the third experiment with 64 subjects exchanging this time Test A with Test B. Finally we developed a final experiment based on ambiguous figures and Stroop effect with sixty 68 subjects, using this time ambiguous figures of animals but able to induce a possible semantic conflict (see Figure3). We retain that in this manner we experienced very different conditions in brain neural correlates. Specific neural pathways were engaged in perception of ambiguous geometrical figures respect to ambiguous forms of animals and, finally, ambiguous forms of animals with added an inner semantic conflict of subjects under Stroop effect.

In conclusion, we used different tests of ambiguous percepts. A total of 256 subjects was involved. According to our quantum model as given in (2.1), we admitted that an individual can potentially have multiple representations of a given choice situation, but he can attend to only one representation at any given time. Strong and immediate ambiguity as induced in the present cases by tests A and B, would consequently induce the subject to suspend his potential consciousness state to be followed from an actualised or manifest state of his consciousness. All the subjects were selected with about equal distribution of females and males, aged between 19 and 22 years. All had normal or corrected-to-normal vision. All they were divided by random selection into two groups (1) and (2). Group (1) was subjected to test A, while the group (2) was subjected to Test B and soon after (about 800 msec after choice for test B) to test A. In all the cases, to avoid the risk to influence the subjects, the question to be asked by tests was posed in the most neutral form.

Each subject was asked to select A=+ or A=- (respectively B=+ or B=-) on the basis of what he was thinking about the figure at the instant of observation..

Finally, It has been shown that perception and cognition in ambiguous figures is influenced by visual angle (Borsellino et al. 1982). Therefore a constant visual angle $V = 2 arctg(S/2D) = 0.33\, rad$. was used with $S$ object's frontal linear size and $D$ distance from the center of the eyes for all the subjects. Each observer was seated at a table with a monitor and computer, was told to look binocularly at the figure, with no fixation point provided, and with random reversals. The observer was requested to stop by pressing a key at the computer when he was aware of having thought of one percept . For each subject we ascertained, after his answer, that he had direct verification of the existing ambiguity in the figure before of his answering. The ambiguous figures were placed in front of the eyes of the observer at a distance of 60 cm, and illuminated by a lamp of 60 W located above and behind the observer's head. The experimental room was kept under daylight illumination. The constant visual angle was realized for each subject using an S object's frontal linear size of about 26 cm for the figure on the monitor.

We also performed a statistical analysis of the results. As previously said, we examined perception-cognition during observation of an ambiguous figure using Tests A and B and thus having two variables that in our approach represent two dichotomous quantum observables $A=\pm$ and $B=\pm$. Quantum observables A and B that attend to the case of geometrical figures cannot be confused with quantum observables that attend to the case of forms of ambiguous animals or to the case of Stroop effect owing to the different neural correlates that each time are involved. Therefore, it resulted appropriate to use a non-parametric test as chi-square, analysing each time singly the results of the experiment under consideration. In brief, for each experiment we evaluated by chi-square rejection or not of the null- hypothesis $H_0$. Probabilities obtained by Test A were considered as to probabilities obtained by Test A/B.

Results

The results are reported in Table n. l. They confirm the presence of quantum like interference that in our cases is ascertained with a statistical significance that overcomes 95%. In Table 1 we give values of $\cos\vartheta(\pm), \vartheta(\pm),$ and of mental wave functions whose meaning was discussed in the previous sections. Consequently, we have a strong evidence that mental states follow quantum mechanics during perception and cognition of ambiguous figures. Using tests, A and B, of Figures 1, 2 and 3, utilizing accurate procedures of experimentation and statistics, we reach a robust conclusion on this subject. On the basis of the (3.1) we confirm, as just said, that mental states follow quantum mechanics during perception and cognitive performance of human brain for ambiguous percepts and cognition. We also find that quantum interference is evidenced when we use tests based on Stroop effect. This is to say, when a semantic conflict is induced during perception. This represents a further strong indication on the possible quantum like behaviour of our mind.

Discussion

The results of the present paper establish that we have quantum like interference during perception and cognition in humans of ambiguous figures and also in the condition of semantic conflict. The presence of quantum like interference, indicates that quantum mechanics has a role in the dynamics of mental states. Cognitive systems are fundamentally quantum like and the quantum representation might serve as the basis for quantum decision making by such cognitive systems.

In detail, our perception-cognitive experiment evidenced that cognitive systems can behave in the quantum like way producing nonzero coefficients of interference. The contexts (corresponding to ambiguity figures) used in this cognitive experiment produced the coefficients of interference that provided a numerical measure of the incompatibility of these contexts. In brief, the experiments outlined deviations of cognitive statistics from classical statistics demonstrated as in classical and quantum wave mechanics.

As consequence, a cognitive system represents a mental context, underlying decision making by a mental wave function, probabilistic amplitude as given in the (2.2) and the (2.3) and explicitly in Table1. Thus, instead of operating with probabilities and analysing (even unconsciously) probabilities of various alternatives, the brain works directly with mental wave functions (probabilistic amplitudes). We conclude that at least some perceptive- cognitive systems have such quantum-like abilities. The brain should result to emulate quantum dynamics at least under some conditions. Such an emulation of quantum dynamics would allow for a three-valued logic in human cognition: true, false and the superposition of true and false. This could explain the peculiar human ability to hold contradictory notions in mind simultaneously, although usually there is collapse to one state or the other. But this ability to see things from "opposite" views might have been valuable in the development of sociability, empathy and even cognitive innovation which seems to depend on seeing things in a radically different way as compared to social or cultural norms.

| Table 1 | Experiment n. 1 | Experiment n. 2 | Experiment n. 3 | Experiment n. 4 |
|---|---|---|---|---|
| p(A=+) | 0.6667 | 0.6154 | 0.5000 | 0.6471 |
| p(A= -) | 0.3333 | 0.3846 | 0.5000 | 0.3529 |
| p(B=+) | 0.5556 | 0.6154 | 0.6250 | 0.3529 |
| p(B= -) | 0.4444 | 0.3846 | 0.3750 | 0.6471 |
| p(A=+/B=+) | 0.6000 | 0.7500 | 0.8000 | 0.1667 |
| p(A=+/B= -) | 0.3750 | 0.8000 | 0.3333 | 0.6364 |
| p(A=-/B=+) | 0.4000 | 0.2500 | 0.2000 | 0.8333 |
| p(A=-/B= -) | 0.6250 | 0.2000 | 0.6667 | 0.3636 |
| p*(A+)=p(B=+)p(A=+/B=+)+p(B=-)p(A=+/B=-) | 0.5000 | 0.7692 | 0.6250 | 0.4706 |
| p*(A-)=p(B=+)p(A=-/B=+)+p(B=-)p(A=-/B=-) | 0.5000 | 0.2308 | 0.3750 | 0.5294 |
| $\Delta$p (+) | 0.1667 | -0.1538 | -0.1250 | 0.1765 |
| $\Delta$p (-) | -0.1667 | 0.1538 | 0.1250 | -0.1765 |
| cos $\theta$ (+) | 0.3536 | -0.2041 | -0.2500 | 0.5669 |
| $\theta$ (+) - value | 1.2094 | 1.7764 | 1.8235 | 0.9680 |
| cos $\theta$ (-) | -0.3354 | 0.7071 | 0.3536 | -0.3354 |
| $\theta$ (-) - value | 1.9128 | 0.7854 | 1.2094 | 1.9128 |
| $\varphi$(+) - mental state wavefunction | $0.577+0.408e^{i\theta (+)}$ | $0.679 + 0.554e^{i\theta (+)}$ | $0.707 + 0.353e^{i\theta (+)}$ | $0.242 + 0.641e^{i\theta (+)}$ |
| $\varphi$ (-) - mental state wavefunction | $0.471+0.527e^{i\theta (-)}$ | $0.392 + 0.277e^{i\theta (-)}$ | $0.353 + 0.500e^{i\theta (-)}$ | $0.542 + 0.485e^{i\theta (-)}$ |
| **Statistical Analysis** | | | | |
| $\chi^2$ value | 5.7143 | 5.5521 | 3.4380 | 6.5750 |
| Statistical significance | $\alpha$ = 0.0168 (*); df=1 | $\alpha$ = 0.0184 (*); df=1 | $\alpha$ = 0.0637 df=1 | $\alpha$ = 0.0103 (*) df=1 |

**Experiment n.1**

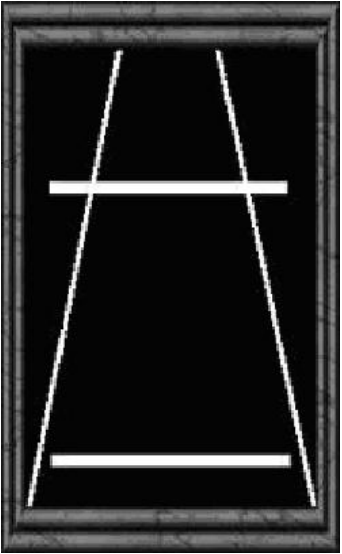
**Test A**

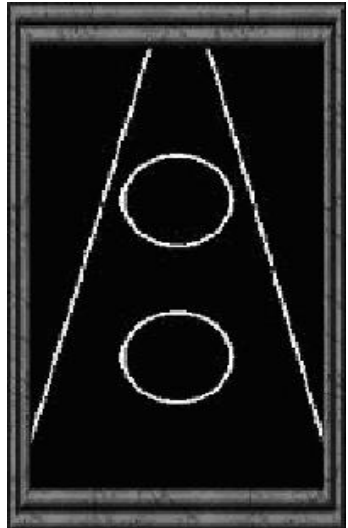
**Test B**

**Experiments n. 2 and n. 3**

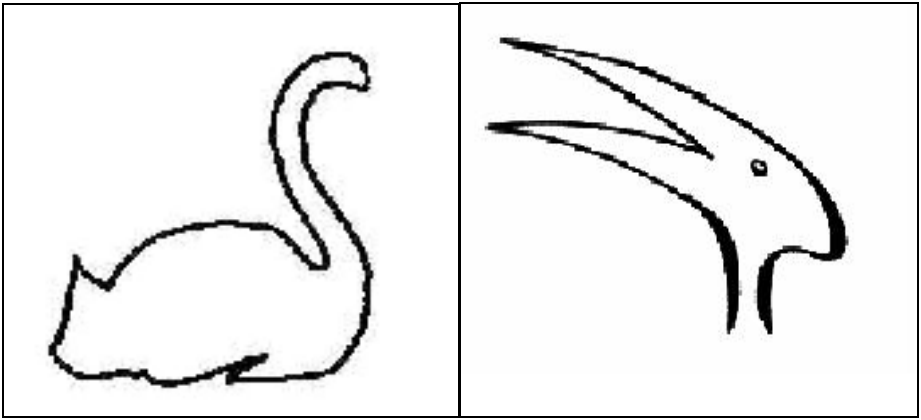

**Test A (Test B)**     **Test B (Test A)**

**Experiment n. 4**

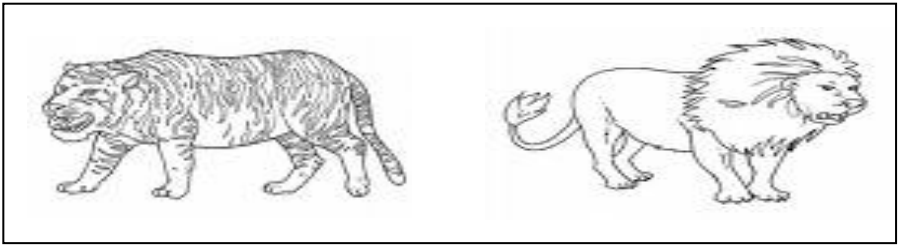

Test A

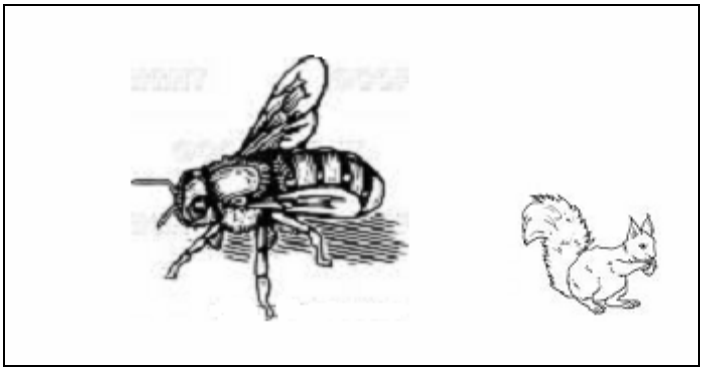

Test B